# Value of Bidirectional V2G Smart Charging Responsive Services: Insights from a Simple CA Model

Pedro M. S. Carvalho and Luís A. F. M. Ferreira

*Abstract* — In this paper, particle-hopping cellular automaton (CA) models of elastic demand are used to investigate the value added to plug-in electric vehicles (PEV) aggregators by adopting vehicle-to-grid (V2G) responsive services. CA models used earlier to study load-sifting responses are modified to capture discharge/ recharge capabilities of V2G. Results on ramping responses from CA are then analysed to discuss the small contribution to system controllability added by V2G responsive services.

*Index Terms* — Ancillary services, Electric vehicles, Demand response, Vehicle-to-grid, Simulation.

## NOMENCLATURE

All variables are discrete-time, indexed by $k = 1, \ldots T$.

- $d_n$: Vehicle $n$ demand output
- $v_n$: Binary decision to shift demand output of vehicle $n$
- $w_n$: Binary decision to discharge the battery of vehicle $n$
- $p$: Aggregate demand of the population of vehicles
- $v$: Aggregate number of shifting decisions in the population of vehicles
- $w$: Aggregate number of discharge decisions in the population of vehicles
- $\pi$: Response price necessary to obtain $v + w$

## I. INTRODUCTION

**V**EHICLE electrification and smart grid technology integration put forward an opportunity for plug-in electric vehicles (PEVs) to provide valuable ancillary grid services [1]. Smart charging (i.e., shifting or reducing PEV demand output during high load periods) will help ensuring that PEVs do not increase peak demand beyond grid's available capacity, while increasing demand-side elasticity and price responsiveness — a valuable service in high renewable supply contexts.

Recently, added services provided by bidirectional vehicle-to-grid (V2G) technologies gained considerable attention from utility commissions and major utilities around the world. The reason seems to stem from the value added by PEV battery discharge capabilities of V2G, which can be enabled as deemed convenient [2]. In this paper, we argue that bidirectional V2G technologies add little controllability to the basic load-shifting smart charging technologies — here referred to as V1G. The argumentation is relevant not just because V2G may be expensive to scale up but also because it may distract utilities from implementing V1G and erode the potential for PEV user engagement based on owners' legitimate concerns around battery capacity decaying [3].

P. M. S. Carvalho and L. A. F. M. Ferreira are with Instituto Superior Técnico, University of Lisbon, and INESC-ID, 1049-001 Lisbon, Portugal (e-mail: pcarvalho@ist.utl.pt; lmf@ist.utl.pt).

## II. LINE OF ARGUMENT

Consider two PEV decision options and the corresponding effects onto the expected aggregate load state: (a) the V1G basic option of shifting demand output of vehicle $n$, $v_n$, and the corresponding effect on the aggregate expected load, $p$; and (b) the V2G option of battery discharging of vehicle $n$, $w_n$, and the corresponding effect on aggregate expected load, $p$. Fig. 1 illustrates the difference between decision options, (a) and (b), and corresponding effects on aggregate expected load.

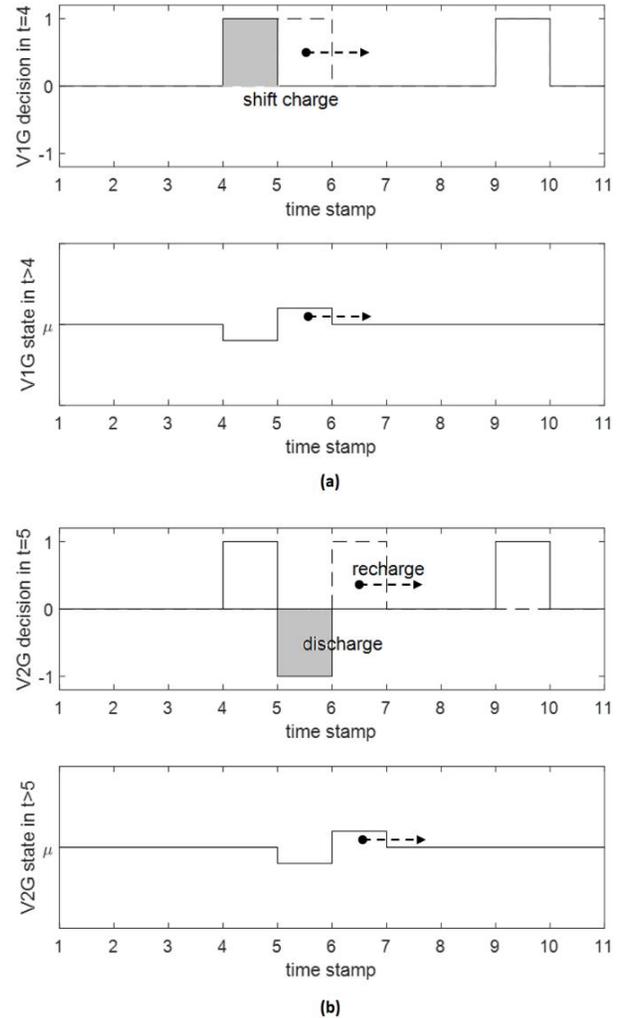

Fig. 1. Two changes in PEV charging schedule, $d_n(k)$, and the corresponding effects on the aggregate load state of the population, $p$: (a) for the V1G decision on shifting charge in time $k = 4$, i.e., $v_n(4) = 1$; and (b) for the V2G decision on discharging the battery in time $k = 5$, i.e., $w_n(5) = 1$.

The argument settles on the similitude between the effects of the different decision options onto aggregate load states and future decision space. By shifting at $k = 4$, the load state is reduced at $k = 4$ and increased at $k = 5$. By shifting, one also gains the opportunity to shift again in a subsequent time, $k \geq 5$ (see Fig. 1 (a) where the arrow denotes such opportunity). If one decides to discharge the battery instead of shifting its charge, one cannot do it without the battery firstly being charged. By discharging it at $k = 5$, the load state is reduced at $k = 5$ but needs to be increased in a subsequent time, $k > 5$, so that the battery is recharged. By assuming that the battery needs to be recharged, one gains the opportunity to choose the time to do so (see Fig. 1 (b) where the dashed box indicates the earlier recharge time and the arrow the opportunity to shift it).

It is apparent that both decisions impact on the aggregate load state by decreasing expected load at the time of decision-making, while increasing it in a subsequent time, both creating an opportunity to shift the charge in the future. From the control space perspective, the main difference is that V2G allows one to decide about when to discharge (after charging), whereas V1G requires one to shift at the time scheduled for charging — as one cannot postpone something already done.

The impact of V2G on PEV aggregators' response can be simulated with CA models of homogeneous load particles, as the ones developed earlier to investigate elastic demand response dynamics [4]. CA models overlook the contextual details of the many possible control implementations [5], but allow focusing on the fundamental differences between V1G and V2G by translating control options into a very simple axiomatic. The model follows.

### III. V2G Cellular Automaton Model

Earlier analysis of demand response dynamics relied upon a simple CA algorithm that simulated real-time control over a population of homogenous load particles. In the algorithm, load particles are indexed by $n = 1, \ldots, N$, where $N$ designates the population cardinality. The aggregate load state $p(k)$ of the population of loads in time period $k$ was defined by:

$$p(k) = \sum_{n=1,N} d_n(k), \quad k = 1, \ldots, T \quad (1)$$

$$d_n(k) \in \{-1, 0, 1\}, n = 1, \ldots, N; \quad k = 1, \ldots, T \quad (2)$$

where $d_n$ represented the demand output of load particle $n$.

Assuming a target output for the aggregate load evolution in time, $p^*(k)$, the following algorithm summarizes the main steps of the CA controller, as implemented:

---

**Cellular Automaton control algorithm**

Set $k = 0; n = 0; v_n(\cdot) = 0 \; \forall n; w_n(\cdot) = 0 \; \forall n$
For $k = 1, \ldots, T$
    For $n = 1, \ldots, N$
        If $p(k) > p^*(k)$
            If $d_n(k) = 1 \land d_n(k+1) = 0$
                Make $d_n(k) = 0$ ; $d_n(k+1) = 1$;
                $v_n(k) = 1$
            Update $p(k)$ with Eq. (1)
        Else Break
Store $p(k)$

---

Assuming V1G decisions to shift charge in time are modeled as ordinary load shifts, i.e., by making $v_n = 1$ in the inner CA condition, then V2G added capabilities to discharge the PEV battery, i.e., the opportunity to make $w_n = 1$, can be modeled by adding a new ELSEIF condition to the inner IF of the CA, as follows (addition in bold):

If $\quad d_n(k) = 1 \land d_n(k+1) = 0$
    Make $d_n(k) = 0$ ; $d_n(k+1) = 1; v_n(k) = 1$
**Elseif** $\quad \mathbf{d_n(k) = 0 \land d_n(k+1) = 0}$
    **Make $\mathbf{d_n(k) = -1}$ ; $\mathbf{d_n(k+1) = 1; w_n(k) = 1}$**
Update $p(k)$ with Eq. (1)

The extended CA algorithm simulates the aggregate state response assuming that there is real-time direct control over PEV decisions. If control is indirect, and each PEV bids a price $\pi_n$ to be rewarded for shifting or discharging, the CA algorithm needs to be changed into an incentive-based controller. To do it, it is sufficient to replace the outer condition on load state,

If $\quad p(k) > p^*(k)$

by a new condition on clearing prices,

If $\quad \pi_n < \pi^*(k)$

where clearing prices are defined by,

$$\pi^*(k) \doteq \pi_n : \sum_{\pi_n < \pi^*(k)} d_n = p^*(k). \quad (3)$$

Refer to [6] for a proof on sufficiency.

### IV. Insights from Simulation

Assuming direct control over PEV or indirect control with perfect information on prices, Fig. 2 illustrates the aggregate state evolution, $p(k)$, obtained for a triangular target, $p^*(k)$, with a population of $N = 5000$. In the subplots, the figure shows the number of V1G and V2G decisions, $v(k)$ and $w(k)$, necessary to follow $p^*(k)$ and the corresponding lattice of PEV responsive actions. The lattices reproduce, for the population of 5000 PEVs, the illustration approach used in Fig. 1 for individual V1G and V2G decisions: $v_n(k)$ and $w_n(k)$ decisions are colored in dark grey for each PEV $n$ after sorting PEVs by bid price, $\pi_n$, in the ordinates of the lattices.

Results show the leading role of V1G over V2G. The number of V2G discharging decisions, $w(k)$, is in orders of magnitude lower than the number of V1G decisions, $v(k)$ — see the second subplot of Fig. 2. Reason is that the capability to ramp-down and to ramp-up aggregate load derive largely from the opportunity to shift in the future. Such opportunity can be created either by shifting a charge (V1G) or by discharging a battery (V2G) — it does not matter much. The important thing is being able to shift such load in the future.

In this context, the only tangible value of V2G is its added capability to create a new opportunity for V1G to shift. But that can be more important than it appears. It may be difficult to find a large enough number of charging actions available to shift in a given time. Thus, the V2G capability to create opportunities for shifting is valuable to accelerate $v(k)$. Fig. 2 shows that V2G actions accumulate at the wave front of V1G actions (lower subplots) to help accelerate the shifting process in order to decrease aggregate demand (first subplot).

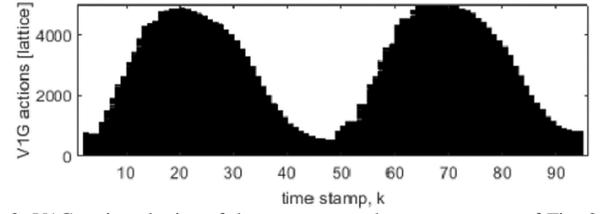

Fig. 3. V1G actions lattice of the response to the same process of Fig. 2 in a situation where V2G is not available.

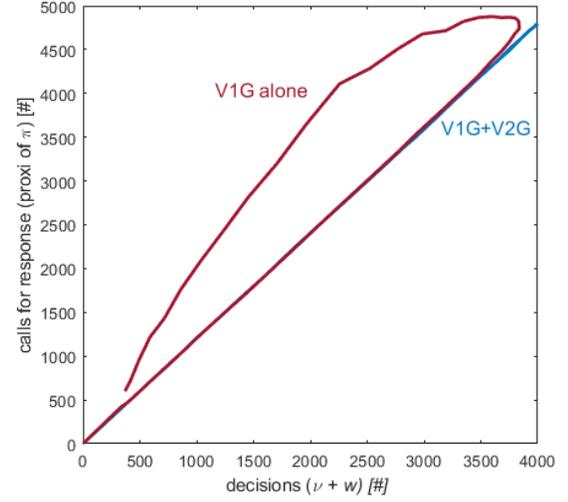

Fig 4. Trajectories of the number of calls for action (a proxy of $\pi^*(k)$) vs. the number of responsive decisions for the process of Fig. 2 in two different situations: with V2G (blue trajectory); and without V2G (red trajectory).

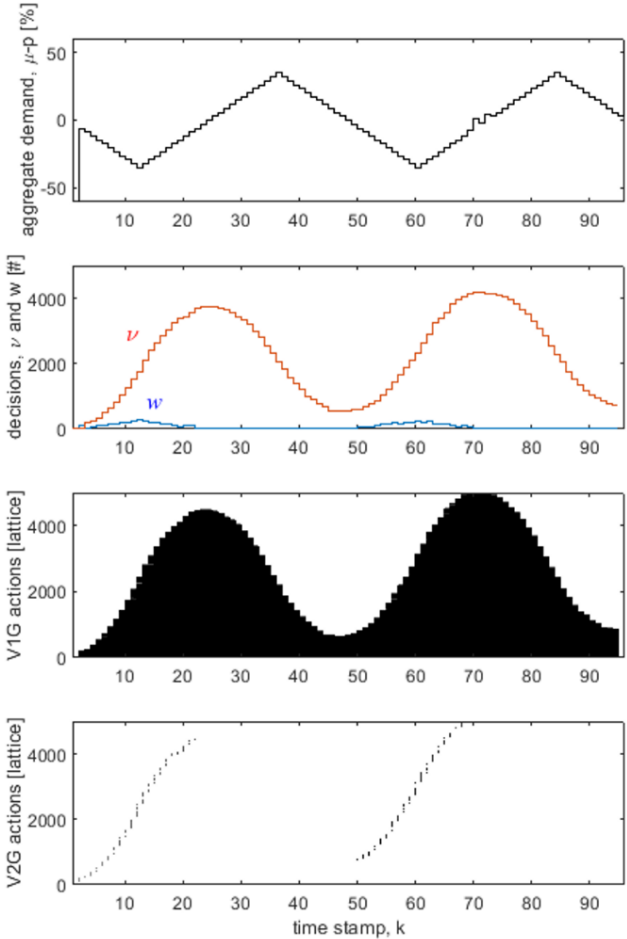

Fig. 2. CA response from a population of 5000 PEV to target triangular wave function with a magnitude of 35% of the expected PEV load $\mu$. The original schedule of charging actions had an average uniform density of 0.167 per period, $k$.

Without V2G, the number of calls for V1G actions would steepen more abruptly in order to accumulate a sufficient number of shifting decisions (cf. Fig 2 V1G actions subplot with Fig. 3). The steepness in the number of calls makes the price rise substantially and, perhaps more importantly, makes response control more difficult when based on prices. This difficulty is illustrated in Fig. 4 by representing the number of calls for action (a proxy of $\pi^*(k)$) against the number of actual response decisions, $v(k) + w(k)$.

Fig. 4 represents trajectories in the space $(v + w, \pi^*)$ in two situations: (*i*) when V2G is not available (red trajectory) and (*ii*) when V2G is available (linear blue trajectory). The V1G alone trajectory exhibits hysteresis, denoting the existence of a strong dynamical effect in the population's response to prices. The dynamics found when ramping aggregate load contribute to increase prices for V1G above prices for V2G solutions. See that one needs to call for more than 4000 shifts to obtain 2500 decisions when ramping-down with V1G alone, and just needs about 3000 calls to get the same responses when V2G is added.

Assuming prices for V1G and V2G are similar, V2G adds value to the aggregator response service. If prices for V2G are higher than for V1G (as expected), then the sole tangible value of V2G seems to be that of making response to prices easier, this way decreasing potential response mismatches.

## V. CONCLUDING REMARKS

In this paper, previously developed CA control algorithms of elastic demand response were used to model V2G and to value its contribution to PEV aggregators that rely on V1G alone. Based on very simple control axiomatic, CA results have shown that V2G is expected to contribute only residually to improve the response capabilities of aggregators. Yet, if response is controlled indirectly by prices or incentives, results have shown that V2G may contribute to simplify control, decreasing response mismatches and ultimately reducing aggregator expenditures.

REFERENCES

[1] FERC, "National Action Plan for Demand Response," *Tech. Rep.*, June 2010.
[2] S. Chhaya, "Distribution System Constrained Vehicle-to-grid Services for Improved Grid Stability and Reliability: Final Project Report," Electric Power Research Institute, California Energy Commission, 2019.
[3] M. Landi and G. Gross, "Battery Management in V2G-Based Aggregations," *Power Systems Computation Conference*, Wroclaw, Poland, Aug. 2014.
[4] P. M. S. Carvalho and L. A. F. M. Ferreira, "Intrinsic Limitations of Load-Shifting Response Dynamics: Preliminary Results from Particle Hopping Models of Homogeneous Density Incompressible Loads," *IET Renewable Power Generation*, vol. 13, no. 7, pp. 1190-1196, 2019.
[5] Y. Zheng, S. Niu, Y. Shang, et al., "Integrating plug-in electric vehicles into power grids: A comprehensive review on power interaction mode, scheduling methodology and mathematical foundation," Renewable and Sustainable Energy Reviews, vol. 112, pp. 424-439, Sep. 2019.
[6] P. M. S. Carvalho, J. D. S. Peres, L. A. F. M. Ferreira et al., "Incentive-Based Load Shifting Dynamics and Aggregators Response Predictability," *Electric Power Systems Research*, vol. 189, pp. 106744, 2020.